# Hybrid Content Distribution Network with a P2P based Streaming Protocol

***Saumay Pushp[1], Dr. Priya Ranjan [2]***

*1. Department of Computer Science and Engineering, Kanpur Institute of Technology, Kanpur, UP-208001, India*
*saumaypushp@gmail.com*

*2. ACES Labs, Department of Electrical Engineering, Indian Institute of Technology, Kanpur, UP-208016, India*
*ranjanp@iitk.ac.in*

***Abstract*— Multicast has been used as a one-to-many approach to deliver information; it is based on the idea that if one packet of data should be transmitted to several recipients, the information should be sent by the origin just one time. In this paper, we propose the use of IP based Pragmatic General Multicast (PGM) [7] to distribute content and to make distribution more efficient, we combine it with a P2P approach. We focus on the problem of data redundancy (at each node), congestion [8] and contention [9] and show how severely it impacts the network economics and the experience of end-user and hence leads to low traffic load and redundancy**

***Keywords*— Multicast, Peer-to-Peer, congestion, contention**

## I. Introduction

Since 20 years, internet has seen an exponential increase in its growth. With more and more people using it, efficient data delivery over the internet has become a key issue. Peer-to-peer (P2P)/efficient data sharing based networks have several desirable features for content distribution, such as low costs, scalability, and fault tolerance. While the invention of each of such specialized systems has improved the user experience, some fundamental shortcomings of these systems have often been neglected. These shortcomings of content distribution systems have become severe bottlenecks in scalability of the internet. The need to scale content delivery systems has been continuously felt and has led to development of thousand-node clusters, global-scale content delivery networks, and more recently, self-managing peer-to-peer structures. These content delivery mechanisms have changed the nature of Internet content delivery and traffic. Therefore, to exploit full potential of the modern Internet, there is a requirement for a detailed understanding of these new mechanisms and the data they serve. The objective here is to propose a protocol which would leverage bit-torrent protocol and IP-multicast.

**The Problem which we are tackling here are:**

**The Problem Of data Redundancy:**

Now imagine the scenario where the number of interested clients increases from few to say around a few hundreds. This is common in case of new files (like movies) getting hosted on websites or critical security patches being made available by software companies. In that case, too much of server bandwidth and bandwidth of access routers is wasted .This leads to each client getting low download rates and bad user experience. We call this problem as the problem of data redundancy and focus to solve it by our hybrid P2P-content distribution system (here Hybrid P2P is used as we augment the Bit-Torrent protocol with our system)

**The Problem Of higher latency and overhead:**

For Small files, Bit-Torrent tends to show higher latency and overhead.

**The Problem of Rapid peer selection:**

Even though several downloaders' might be physically close to each other and downloading the same file (e.g. several clients on a LAN downloading a software patch), the tracker returns a random list of peers to which a new downloader should connect to. This leads to wastage of resources because of redundant downloads of same pieces by peers close to each other.

**Background and Related Work**

### *A. P2P (Peer to Peer)*

Several years ago, the use of centralized services was the target of several research lines. Its simplicity provided developers and researchers a straightforward model to accomplish several tasks such as sharing files, sharing resources, store data, among others. The main two problems with this approach are **1) Failure tolerance**, since there is a single point of failure centralized services may not provide a proper degree of availability. **2) Scalability**, as the number of users trying to access a centralized service grows bottlenecks may be present as well as denial of service. To overcome these issues, distributed systems such as P2P (Peer-to-peer) overlay networks were created. Indeed, the use of these systems is so high that up to 60% of the Internet traffic is attributed to P2P systems. In a P2P system, peers collaborate to form a distributed system for the purpose of exchanging the content. Peers that connect to the system behave both as servers and clients. A file that one peer downloads is often made available for upload at other peers. The participation is purely voluntary. However, a recent study [2] has shown that most content-serving hosts run by end-users suffer from low availability, and have relatively low capacity network connections (modem, cable modems, or DSL routers). Users interact with a P2P system in **two ways:** they attempt to locate objects of interest by issuing search queries, and once relevant objects have been located, users issue download requests for the content. P2P systems differ in how downloads proceed, once an object of interest has been located. Most systems transfer content over a direct connection between the object provider and the object seeker. A latency-improving optimization in some systems is to download multiple object fragments in parallel from multiple replicas. A recent study [3] has found the peer-to-peer traffic of a small ISP to be highly repetitive, exhibiting great potential for caching and other techniques for enhancing performance.

### *B. Bit-Torrent Protocol and Algorithms*

Bit-Torrent is a Peer-to-Peer file sharing protocol, originally designed and implemented by Bram Cohen [1]. With Bit-Torrent based file download, when multiple clients are downloading the same file at the same time, they upload pieces of the file to each other, redistributing the cost of upload to downloader's. Thus, Bit-Torrent makes hosting a file with potentially unlimited number of downloader's affordable. To start a Bit-Torrent deployment, a static file with a .torrent extension, accessible to all downloader's is placed on an ordinary server. The torrent file contains file related information like the length of the file, name, hashing information and the URL of the tracker. Trackers help downloader's find each other, while speaking a simple protocol layered on top of HTTP. A downloader on initialization **contacts the tracker** sending information like name of the file it is downloading and the **port on which it is listening**. The tracker then responds with a list of peers which are also downloading the same file. Based on this, the downloader's then connect to each other. To start the download, one of the downloader's which already has the complete file (**called seeder)** must be started. **The role of tracker** is essentially limited to assisting peers finding each other and keeping statistics and thus the load on it is minimal. In-fact, even if the tracker goes offline after all the downloader's have started, the protocol is not severely affected. In order to keep track of which peers have what, Bit-Torrent cuts files into pieces of fixed size, typically a quarter megabyte **(**typically **250 Kb** each**)**. Each downloader reports to all of its peers what pieces it has. To verify data integrity, the SHA1 hashes of all the pieces are included in the .torrent file, and peers don't report that they have a piece until they've checked the hash. Peers continuously download pieces from all peers which they can. They of course cannot download from peers they aren't connected to, and sometimes peers don't have any pieces they want or won't currently let them download. Selecting pieces to download in a good order is important for having good performance. For example, poor piece selection criteria can result in all peers downloading the same set of pieces and thus may end up with none of them having any piece to upload to the other. Bit-Torrent follows a strict priority order in which once a single sub-piece has been requested, the remaining sub-pieces from that particular piece are requested first before sub-pieces of other pieces. Several piece selection criteria have been suggested, including the following:

- **Rarest First:** Following this policy, the peers download pieces which are rarest amongst their own peers. It ensures downloader's to have pieces which their peers would want to be uploaded. The pieces which are generally available amongst the peers are left for later download so that the likelihood that a peer that is currently offering the upload will later not have anything of interest is reduced.

- **Random First:** When the downloading starts, the peers have nothing to upload, thus it is important to get a complete piece as quickly as possible. The pieces to be downloaded are selected at random until the first complete piece could be assembled. After that the strategy changes to rarest first. The peers are responsible for maximizing their own download rates. The peers do this by downloading from whichever peer they can and deciding which peers to upload to via a variant of tit-for-tat. To cooperate, the peers upload, and to not cooperate they choke peers. **Bit-Torrent choking algorithm** attempts to achieve **Pareto efficiency** [17] by having the peers reciprocate uploading to the other peers which upload to them. Unutilized connections are also uploaded to on a trial basis to see if better transfer rates could be found using them.

**Some Drawbacks of the Bit-Torrent protocol:**

• For Small files, Bit-Torrent tends to show higher latency and overhead.
• Even though several downloader's might be physically close to each other and downloading the same file (e.g. several clients on a LAN downloading a software patch), the tracker returns a random list of peers to which a new downloader should connect to. This leads to wastage of resources because of redundant downloads of same pieces by peers close to each other.

### C. *Bit-Torrent Location-aware Protocol*

As mentioned above, the original Bit-Torrent protocol can lead to peers geographically distant from one another exchanging data when peers close by are also present, leading to suboptimal performance. A location-aware Bit-Torrent protocol has been proposed in [19]. However, the proposal is in a very lose form with no real world implementation or performance results. It requires each Bit-Torrent client to supply its approximate geographical location (longitude and latitude) when contacting the tracker to get the peer list. The tracker knows geographical locations of all downloader's and thus returns the list of peers to the original requester which are closer to it, instead of returning a random list (as in case of the original Bit-Torrent tracker).Several issues arise here. Firstly, this protocol is not compatible with the original Bit-Torrent protocol and requires changes at the trackers. Secondly, assuming that the geographical location of a client would be known is not realistic. Thirdly, clients located close to each other geographically may not be having a fast network link between them and might be separated by several hops in terms of routing. Finally, absence of any implementation of this protocol makes one skeptical about the relative performance gain of it.

### D. *The Problem of Data Redundancy*

Consider the scenario in the given figure. Each client establishes an independent TCP connection to the file server to fetch the file. If all the clients need to download the file at the same time, nine parallel TCP connections with file server as the source have to be started. This means that the server opens 9 different sockets to serve each TCP connection and essentially transmits the same data through each of these sockets. Thus, nine exact copies of the file available at server are sent across the link connecting the file server and the core router. The core router in turn sends 3 copies of the same data on each of the access links. Now imagine the scenario where the number of interested clients increases from nine to say around a few hundreds. This is common in case of new files (like movies) getting hosted on websites or critical security patches being made available by software companies. In that case, too much of server bandwidth and bandwidth of access routers is wasted. **As a consequence, network congestion and contention** occurs and an incremental increase in offered load leads either only to small increase in network throughput, or to an actual reduction in network throughput.

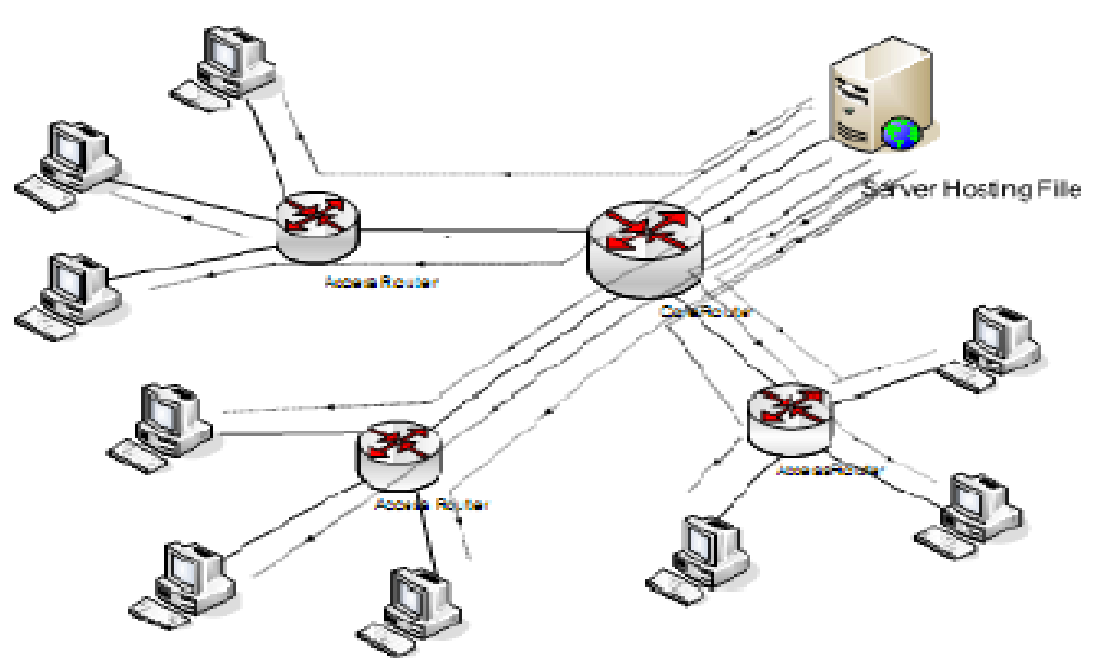

Fig1. Scenario describing Data distribution

This leads to each client getting low download rates and bad user experience. We call this problem as the problem of data redundancy and work towards solving this by multi-casting.

## II. Performance Study of Content Distribution Models

Earlier, we talked about the content distribution models, including the Peer-to-Peer Systems model. With the help of an example scenario, we also illustrated the problem of same data being re-transmitted over internet links, leading to degraded performance and higher running costs. Now, we present the results of a large scale experimental study to understand the performance of each of the content distribution models. The study was conducted using the **Emulab** [12] emulation facility.

### A. *Experimental Setup*

- **Network Topology**

  The first step towards performing experiments on Emulab is to specify the network topology and the specification of hardware and software on each node of the network. This is done with the help of a topology specification script written in tcl programming language, in a format identical to that of NS2 [21].Internet can be assumed to be composed of two entities:

  **Backbone Network:** It consists of the high bandwidth, high delay, and long distance network links, which typically run across continents and countries. These backbone links are generally hosted by various Internet Service Providers (ISPs) and account for the main cost in running the internet.

  **High Speed LANs:** Most organizations today have access to high speed local area networks (LANs) which in turn are connected to the backbone internet via particular nodes (routers). Such LANs are generally error-free and congestion free and are administered by the local organizations. Since the major cost in running Internet is in maintaining the backbone network, the ISPs are generally concerned about transferring the data across backbone links in the most cost-effective manner. The cost for a link is proportional to the amount of data (or the number of bytes) transferred across the link. In this study, we try to understand the typical amount of traffic which the ISPs need to transfer to support the different content distribution models. Also, as we show in this study, most of the current models end-up sending the same data again and again over the same links. We are interested in designing a **hybrid CDN structure** which restricts such retransmissions.

  Fig 2 illustrates the network topology used for this performance study on Emulab.

The internet backbone is made up of four core routers, named coreRouter0, coreRouter1, coreRouter2 and coreRouter3. Each of the core Routers run on the Red Hat Linux 9.0 Standard operating system. The four core routers are all connected to each other in a symmetrical manner and thus there are total six core links named corelink0 ... corelink5. Each of the core links is a 10Mb link with a 20 ms end-to-end delay and a Drop Tail queue. Three of the core routers (coreRouter0, coreRouter1 and coreRouter2) are each connected to a set of three high speed LANs via routers (router0, router1 and router2). Each of the three routers runs the Red Hat Linux 9.0 version of operating system. The link between a router and a core router is a 2Mb link with a 10 ms end-to-end delay and a Drop Tail queue. Each router is in turn connected to three 10 Mbps LANs (for example, router0 is connected to lan0, lan1 and lan2). Each LAN is composed of 4 end nodes and a switch. The nodes are named from node0 to node 35 (total 36 end-nodes/clients). A dedicated node (named seeder) is connected to coreRouter3 via a 2Mb link.

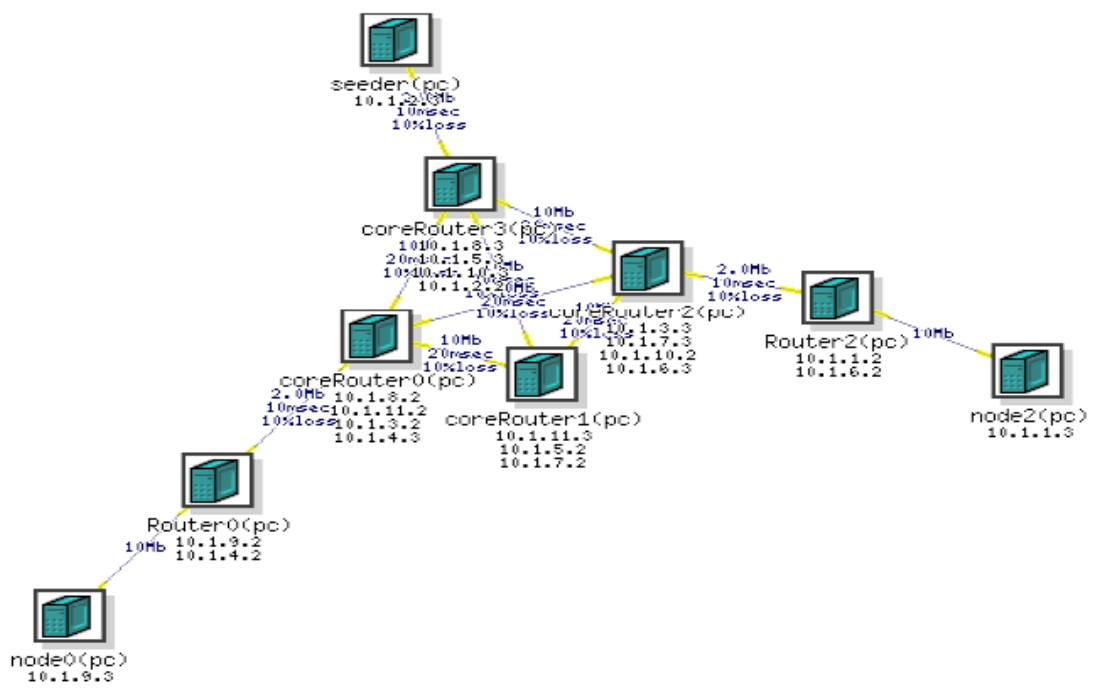

Fig 2. The experimental setup used for the performance study

- **Performance Metrics**

In this study, we are concerned about quantifying the amount of data transmitted over backbone links in the various content distribution models. Thus, we measure **two** key metrics in each experiment run, for each link, in each direction:

**Number of Bytes:** This represents the raw amount of data transferred over a link in a particular direction.

**Stress:** This represents the ratio of number of total packets transmitted over a link and the number of unique packets transmitted over the link. For example, a stress of **2** represents a case where each packet is transferred twice over a link. As mentioned earlier, the running cost of a link for the ISP is proportional to the raw amount of data transferred over a link. A higher link stress refers to the case where higher redundant transmissions of the same data are happening over the link, thus wasting the bandwidth.

**Emulab** has simple support for tracing and monitoring links and LANs. For example, to trace a link:

**set link0 [$ns duplex-link $nodeB $nodeA 30Mb 50ms DropTail]**
**$link0 trace**

The default mode for tracing a link (or a LAN) is to capture just the packet headers (first 64 bytes of the packet) and store them to a tcpdump output file. In addition to capturing just the packet headers, one may also capture the entire packet:

**$link0 trace packet**

By default, all packets traversing the link are captured by the tracing agent. To narrow the scope of the packets that are captured, one may supply any valid tcpdump style expression:

**$link0 trace monitor "icmp or tcp"**

One may also set the **snaplen** for a link or lan, which sets the number of bytes that will be captured by each of the trace agents:

**$link0 trace_snaplen 128**

In our experiments, we set the snaplen to 1600 bytes.For each link **(**say link0, between nodeA and nodeB**)**, 2 trace files of interest are generated by tcpdump: **trace nodeAlink0. recv** and **trace nodeB-link0.recv.** Here, the first trace file stores the packets sent by nodeA to nodeB over link0, while the second file stores the packets sent by nodeB to nodeA over link0. To analyse the tcpdump trace files, we modified a well known tool tcptrace . We added a module in the tcptrace code to calculate the **MD5 checksum** of payload of each tcp packet and store the checksums of all payloads in a file. The number of checksums is equal to the total number of packets transmitted over a link. We **then calculate the number of unique checksums in the file**, which represents the number of unique packets transmitted. The ratio of these two gives the link stress. Also, the total number of bytes from payloads of all tcp packets on a link can be easily calculated from tcptrace.

In our experiment we designed a Bit-Torrent client which supports the following

• Must support a console based interface to allow remote execution over Emulab nodes

• We preferred it to be in java so that Datagram sockets could be used to extend it to support IP multicast [15]

### *B. Performance Evaluation*

**Peer-to-Peer Model:**

Table below shows the link statistics for the file download using Bit-Torrent on each of the end/client nodes.

| Link | Direction | No. of Bytes | Stress |
|---|---|---|---|
| coreLink0 | coreRouter0 − > coreRouter1 | 4.1 MB | 5.712 |
| coreLink0 | coreRouter1 − > coreRouter0 | 3.6 MB | 6.078 |
| coreLink1 | coreRouter2 − > coreRouter1 | 3.9 MB | 5.429 |
| coreLink1 | coreRouter1 − > coreRouter2 | 3.1 MB | 5.896 |
| coreLink2 | coreRouter0 − > coreRouter2 | 2.8 MB | 6.368 |
| coreLink2 | coreRouter2 − > coreRouter0 | 4.0 MB | 5.221 |
| coreLink3 | coreRouter0 − > coreRouter3 | 15 KB | 2.800 |
| coreLink3 | coreRouter3 − > coreRouter0 | 0.8 MB | 1.544 |
| coreLink4 | coreRouter1 − > coreRouter3 | 16 KB | 3.620 |
| coreLink4 | coreRouter3 − > coreRouter1 | 1.2 MB | 1.855 |
| coreLink5 | coreRouter2 − > coreRouter3 | 15 KB | 2.880 |
| coreLink5 | coreRouter3 − > coreRouter2 | 0.9 MB | 1.366 |
| link0 | coreRouter0 − > router0 | 8.5 MB | 6.630 |
| link0 | router0 − > coreRouter0 | 6.9 MB | 7.013 |
| link1 | coreRouter1 − > router1 | 9.3 MB | 7.155 |
| link1 | router1 − > coreRouter1 | 6.7 MB | 7.300 |
| link2 | coreRouter2 − > router2 | 6.8 MB | 6.516 |
| link2 | router2 − > coreRouter2 | 8.0 MB | 6.933 |
| link3 | coreRouter3 − > seeder | 47 KB | 3.184 |
| link3 | seeder − > coreRouter3 | 3.0 MB | 2.797 |

In the P2P model, clients download the file in a collaborative manner. Instead of depending only on the seeder for the file download, each client fetches data packets from other clients as well, which may also be downloading the same file. Thus, in this case, clients have TCP connections between them, in addition to TCP connections with the seeder. Unlike the WWW model where the uplink capacity of most links remain largely un-utilized, in P2P model, since clients are also responsible for uploading packets to other clients, thus the uplink capacity is also used in P2P model. The other important observation is regarding the link stress. We observe that link stress values are smaller in case of the core links. This means that there are lesser number of duplicate packet transmissions happening over the internet links, thus avoiding the wastage of resources. This is due to the fact that each client observes the data pieces which are available with other clients and fetches them as well, instead of fetching pieces always from the seeder.

**WWW Model:**

In the WWW model, each client downloads the file from the server (seeder in this case) independently of each other (using wget [13]), by establishing independent HTTP connections over TCP with the server. Each HTTP connection is composed of two independent TCP connections: (1) From client to server to send request and acknowledgement packets (2)

| Link | Direction | No. of Bytes | Stress |
|---|---|---|---|
| coreLink0 | coreRouter0 − > coreRouter1 | 0 | 0 |
| coreLink0 | coreRouter1 − > coreRouter0 | 0 | 0 |
| coreLink1 | coreRouter2 − > coreRouter1 | 0 | 0 |
| coreLink1 | coreRouter1 − > coreRouter2 | 0 | 0 |
| coreLink2 | coreRouter0 − > coreRouter2 | 0 | 0 |
| coreLink2 | coreRouter2 − > coreRouter0 | 0 | 0 |
| coreLink3 | coreRouter0 − > coreRouter3 | 1.3 KB | 12.000 |
| coreLink3 | coreRouter3 − > coreRouter0 | 14.5 MB | 11.585 |
| coreLink4 | coreRouter1 − > coreRouter3 | 1.3 KB | 12.000 |
| coreLink4 | coreRouter3 − > coreRouter1 | 14.5 MB | 4.573 |
| coreLink5 | coreRouter2 − > coreRouter3 | 1.3 KB | 13.000 |
| coreLink5 | coreRouter3 − > coreRouter2 | 14.5 MB | 4.485 |
| link0 | coreRouter0 − > router0 | 14.5 MB | 11.585 |
| link0 | router0 − > coreRouter0 | 1.3 KB | 12.000 |
| link1 | coreRouter1 − > router1 | 14.5 MB | 4.573 |
| link1 | router1 − > coreRouter1 | 1.3 KB | 12.000 |
| link2 | coreRouter2 − > router2 | 14.5 MB | 4.485 |
| link2 | router2 − > coreRouter2 | 1.3 KB | 13.000 |
| link3 | coreRouter3 − > seeder | 3.9 KB | 37.000 |
| link3 | seeder − > coreRouter3 | 43.6 MB | 8.315 |

From server to client to send the data packets. Intuitively, the bandwidth utilization in the uplink direction (from client to server) is much lower as compared to the downlink direction (from server to client).Table 3.1 shows that corelink0, corelink1 and corelink2 are not utilized for any bytes transfer. This is due to static routing path between the seeder and the clients, which does not include these 3 links but uses other 3 parallel and equivalent (in terms of bandwidth and delay) links: corelink3, corelink4 and corelink5. Table 3.1 also illustrates that the amount of data transferred in the uplink direction is much lower than that transferred in the downlink direction. The most important result shown in Table 3.1 is the number of bytes transferred on each link in the downlink direction. For example, on coreLink3, 14.5 MB of data is transferred downlink. This is explained by the fact that coreLink3 connects seeder with coreRouter0, which in turn connects to all clients on lan0, lan1 and lan2. Since each of the 12 clients on lan0, lan1 and lan2 download the data independently; the shared file of 1MB is downloaded 12 times through coreLink3. The extra bytes transferred correspond to other overheads of HTTP. Another interesting result shown in Table is the stress on each of the internet links. We see that a stress of around 4 is common on most links in the downlink direction. Note that stress is not necessarily equal to the ratio of total bytes transferred over a link and the number of bytes in the download le. This is because for calculating link stress, we calculate MD5 checksum of each TCP packet payload, which contains HTTP headers, etc. For packets which have similar data contents, these HTTP headers may slightly differ, for example only by timestamps. Due to this, the MD5 checksums of similar packet payloads differ, thus reducing the overall stress. A potential way of dealing with this problem is the use of another hash function to calculate the checksum which is suitable for detection of near duplicate contents.

### *A. IP Multicast as a Content Distribution Model:*

IP Multicast is a particularly attractive alternative for content distribution in such scenarios. All the clients can initially send IGMP request messages to join a multicast group and the source can multicast the data on this group. Since routers are aware of the physical topology and positions of clients, the data traverses the shortest path to reach each of the clients, guaranteeing optimal download time. Although such an approach is promising, it is not viable in today's Internet because of lack of support of IP Multicast on Internet. This means that two nodes on the Internet do not necessarily have a route between them which is IP Multicast enabled **There are several reasons why IP Multicast is not available on the Internet**. These include:

• Most routers on the Internet lack support for IP Multicast. Recollect that to support IP Multicast, a router needs to perform several additional operations like duplication of packets with PIM, IGMP support, Multicast forwarding etc. The routers available on Internet simply do not have resources or capabilities to perform all such operations. Upgrading such existing routers is clearly infeasible.

• Congestion control schemes are not well defined for multicast.

• Pricing policies in multicast are not clear. Hence, there are no incentives for the ISPs to be interested in deploying multicast support in the networks.

Therefore, it is almost clear that utilizing IP-level multicast for large scale content distribution in above mentioned scenarios is not feasible. The problem of IP Multicast as an unreliable protocol is that it works over UDP. This means that there is no guarantee that a packet multicast over UDP will be successfully received by other clients. Since IP Multicast does not have any mechanisms for rate control and checking packet losses (due to random errors etc.),it is not necessary that pieces shared by clients would be received by all other clients on the island. The clients which have low received buffer or which are busy with other operations often are unable to completely receive packets sent over multicast. We tackle the above problem in providing more efficient data sharing through the concept of **3-way Hand shake** [22] and propose a method which co-exist with the standard Bit-Torrent protocol and leverage IP Multicast to distribute downloaded pieces to other Bit-Torrent clients on the same network

## III. OBSERVATIONS

Although IP Multicast is not available on the Internet, we have observed that most organizations have it enabled on their local networks. This is so because upgrading a few routers to support IP Multicast on the local networks is relatively an easier task as compared to upgrading millions of routers on the Internet. Besides, problems like absence of congestion and rate control mechanisms for IP Multicast are less severe on local networks which are typically high speed, free from error and congestion. Lastly, pricing policies for use of links within the local network is not very important as these links are hosted by organizations themselves and not by foreign ISP. Based on our observations, we assume that most of such islands are IP Multicast enabled. We observed that bulk data download happens using Peer-to-Peer systems and there are several instances where there exist multiple downloading clients within the same islands (for example, university networks etc.). These clients are unaware of each others presence and fetch data packets from outside the island over Bit-Torrent.

## IV. OUR APPROACH

We used a highly modular approach to the problem. We figured out that there are basically 5 parts to the program:

**1. Database Manager:** This takes care of the list of chunks of different files available on the network.

**2. Chunk Maker/Assembler:** This creates chunks of a file and maintains a mechanism for testing the integrity of each chunk. It also assembles the chunks into a complete file when all the chunks of a file have been downloaded.

**3. Chunk Sender/Receiver:** This communicates on a single port with another host on a defined port and transfers file reliably. This throws back problems if encountered in the process or flags a success message if it is successful.

**4. User Interface:** This is where the user interacts with the program. We have 2 such interfaces, one is a GUI and another is a console one. Here the user can ask to share a file on the network and fetch a file from the network.

**5. The Head:** This interacts with every other part and decides what to do when. It basically deploys the work to other modules and also performs a **3-way handshake** before a communication begins on a defined port using the Chunk Sender/Receiver.

Multicast packets on an island can be lost or delayed due to two things:

**1.** The clients and links on a LAN show abnormal behavior (due to load or miss-configuration) leading to random packet losses.
**2.** There is congestion on the LANs due to other heavy traffic being exchanged by clients, e.g., VoIP etc.

**The 3-way handshake**:
The tracker when requested to fetch a file from the network, it does the following:
**a.** Asks the Managed Hash Table for the information of the locations of the chunks.
**b.** Now for each chunk, it contacts the Tracker of another host sending a Type1 packet requesting a chunk.
**c.** The peer host's tracker sends back a packet which can be:
• Type2 packet: This says that the peer host has accepted the request and it is designating a port for sending the chunk.
• Type3a packet: This says that the peer host does not have the chunk requested and thus is negating the connection.
• Type3b packet: This says that the peer host has the chunks but currently does not have any free ports to take the request.
**d.** If Type3a is received then the tracker tries to request the file from another source, if available.
**e.** If Type3b is received then the tracker would look for other sources and if it runs out of other sources it ask the same host after some time.
**f.** If Type2 packet is received, the Tracker sends a Type4 packet that carries the information about which port of this user would be listening for the packets and starts the Downloader.
**g.** On reception of Type4, the Up-loader is called.
**h.** If on any of these communications, a timeout is faced, it is gracefully handled.

Thus we achieve a 3-way handshake similar to TCP for starting up the chunk transfer.
This kind of a handshake happens for every chunk transfer. We then ask the Downloader/Up-loader to simply transfer the file over a designated port. For every get file request, the Tracker starts transferring chunks in a batch. This ensures that no get file request hogs all the ports available for communication.

## V. PERFORMANCE EVALUATION

### A. *Topology*

For the sake of completeness, the topology is shown in Fig 2.The components of the topology are the end clients (node0 to node2), the access routers (router0 to router2) and the core routers (coreRouter0 to coreRouter2). There are two types of links in this topology:
**Core links:** which serve the traffic across the internet by connecting the core routers; and
**Access links:** which are used to provide internet access to the islands consisting of various high-speed LAN's. Since the two types of links carry different type of traffic, we show the evaluation of both types separately.

Each island in our experimental topology consists of 3 high-speed (10 Mbps) LANs. All the LANs are connected to each other via the access router (i.e., router0, router1 or router2). Each of the access routers runs the Red Hat operating system. In order to allow IP Multicast across different LANs on the same island, we run mrouted [8] on each of the access routers. The mrouted utility is an implementation of the Distance-Vector Multicast Routing Protocol (DVMRP), an earlier version of which is specified in RFC-1075 [10].It maintains topological knowledge via a distance-vector routing protocol (like RIP, described in RFC-1058 [11] ), upon which it implements a multicast datagram forwarding algorithm called Reverse Path Multicasting. The mrouted utility forwards a multicast datagram along a shortest (reverse) path tree rooted at the subnet on which the datagram originates. The multicast delivery tree may be thought of as a broadcast delivery tree that has been pruned back so that it does not extend beyond those sub networks that have members of the destination group. Hence, datagrams are not forwarded along those branches which have no listeners of the multicast group. The IP time-to-live of a multicast datagram can be used to limit the range of multicast datagrams. Thus, any multicast packet in one of the LANs reaches all other LANs on the same island, provided their are clients on the other LANs who have subscribed to the corresponding multicast group .Also, we set the **TTL value** of multicast packets to 3 to allow them to cross multiple levels of multicast enabled routers. Note that a **TTL value** of 1 means that packets are limited to the same subnet.

### B. *Performance Metrics*

**Number of Bytes:** This represents the raw amount of data transferred over a link in a particular direction.
**Stress:** This represents the ratio of number of total packets transmitted over a link and the number of unique packets transmitted over the link.
**Time for Download:** This represents the total time each client takes to download the file.

### C. *Comparison*

The seeder serves a file of size 1 MB. All the results reported in this section have been obtained after proper averaging over 5 to 10 runs of each experiment.

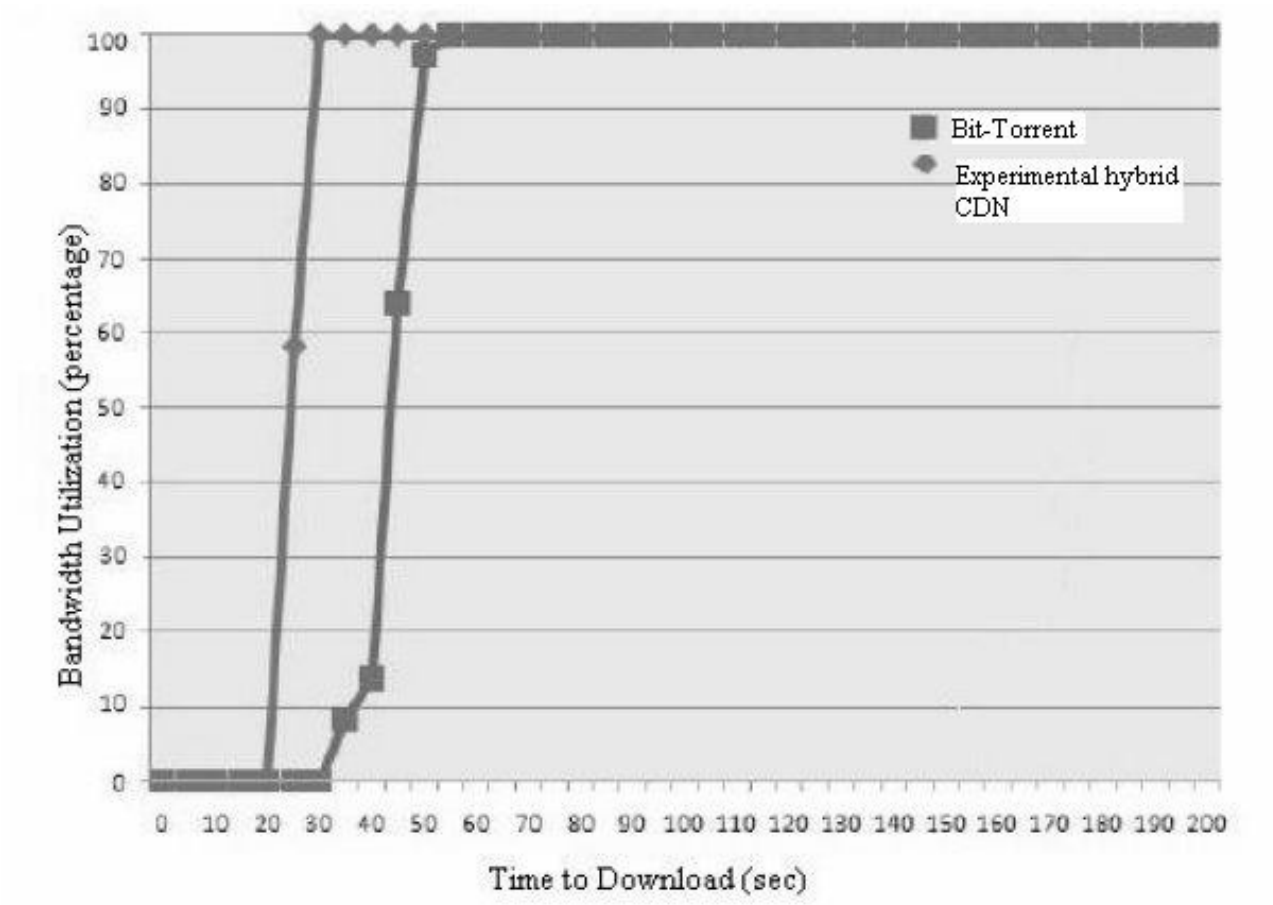

Note that the steeper the plot is, the faster is the completion of download for all the clients. In the above figure 100 % of clients complete their download within 30 seconds while using Hybrid CDN. It takes about 60 seconds for all the nodes to complete their download using Bit-Torrent.

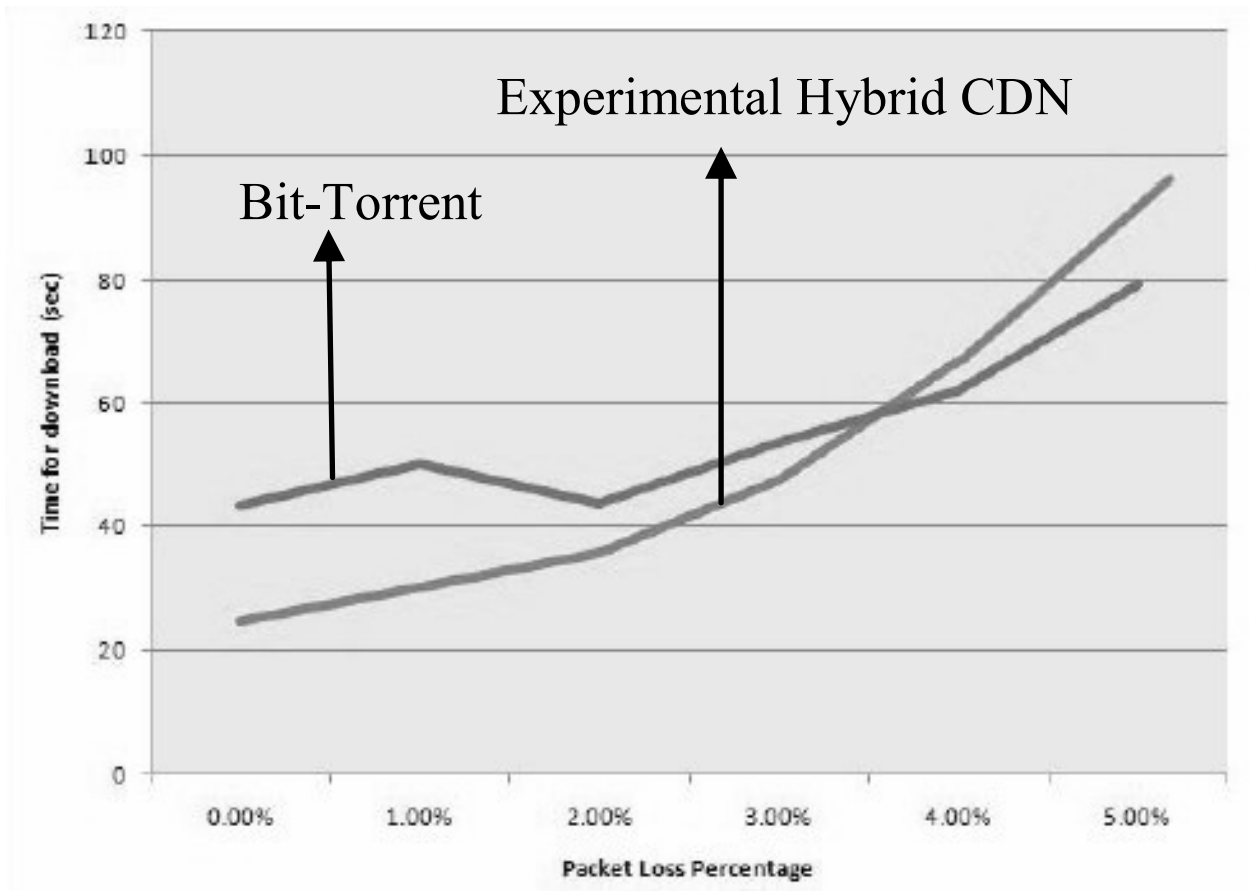

We varied the packet loss percentage from 0-5%. The Experimental Client's download time increase after 4% packet loss, due to the fact that during multicasting maximum of the packets get lost and they are retransmitted using unicasting.

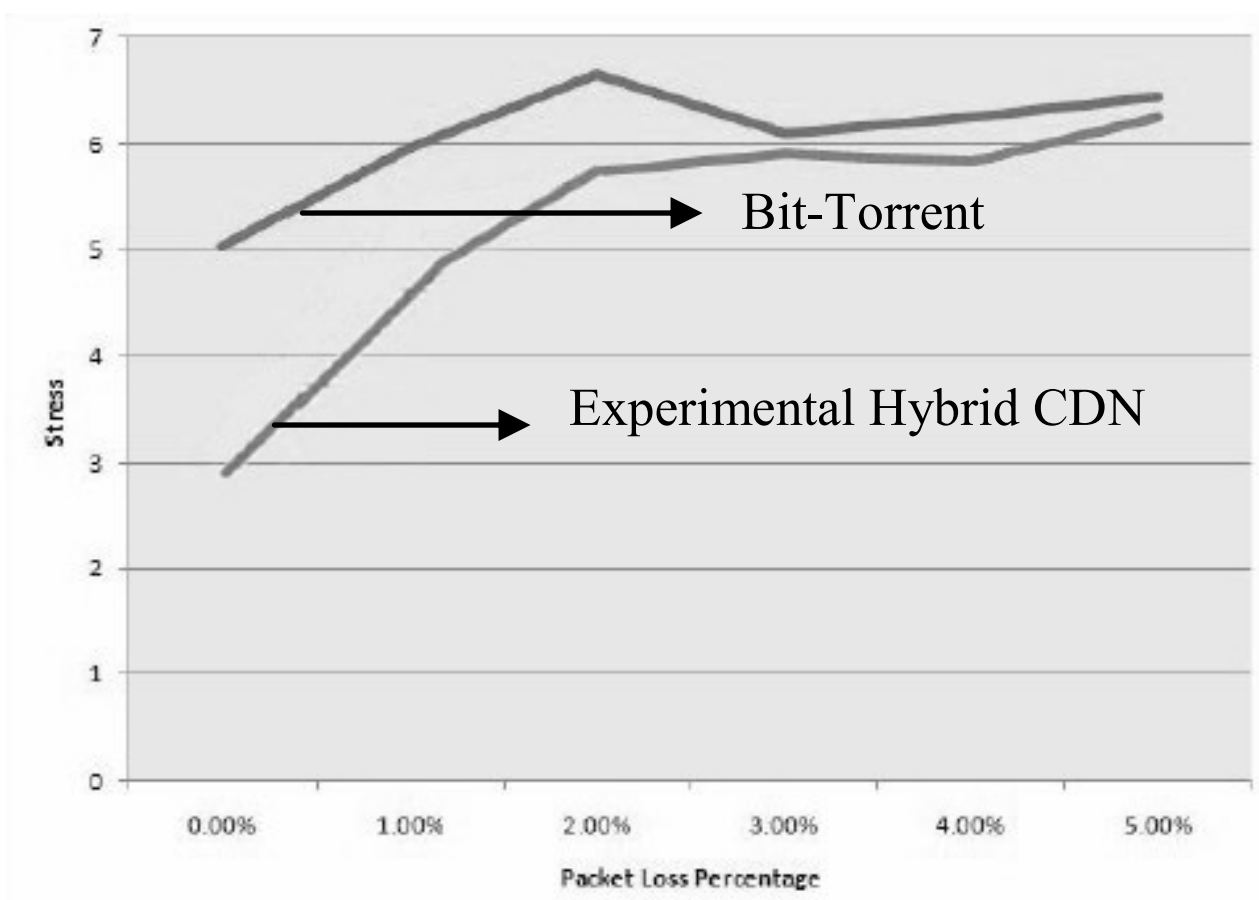

Figure showing Variation in Stress Vs Packet Loss Percentage

In the below Fig we varied the CBR (Constant Bit Rate) from 0-10% to evaluate congestion in the network

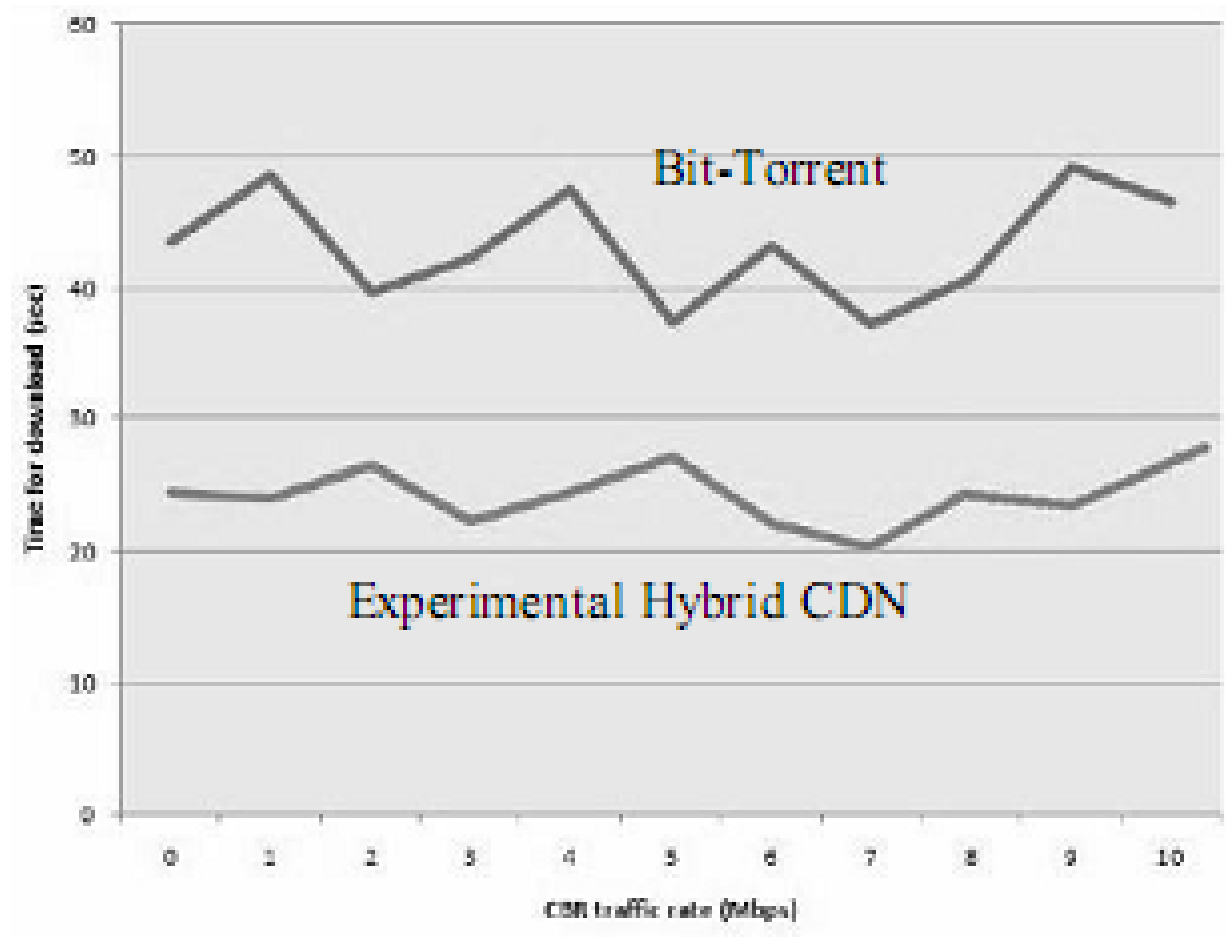

## VI. Conclusion

We obtained the following three important results with the Hybrid CDN Model:

- Reduction in download time of each client using Hybrid CDN by 48% over Bit-Torrent and by 86% over WWW protocol
- Reduction in traffic load on Internet links and ISPs.
- Reduction in the wastage of resources like bandwidth due to redundant packet.

Downloading time is the most critical performance metric for normal Internet users, whose experience with the system is largely determined by how fast they can download file from the Internet. Also, recent applications of Peer-to-Peer systems like distributing the software updates and the images of operating systems, etc., over large networks spread across a geographically distributed area depend heavily on the download time for each computer. The delay in download in [10] can be overcome by the use of Hybrid CDN like model leveraging the IP-Muticast and Bit-torrent protocol applicability. Most ISPs today observe heavy traffic load on their Internet links due to increasing number of users using Peer-to-Peer file sharing systems. Due to competition, ISPs are forced to reduce tariff continuously resulting in reduction in the margins of profit. However, with more users migrating to a system like Hybrid CDN, the load on ISP resources (Internet links) can be reduced by as much as 65%, for the comparable amount of downloads by end clients. Thus, the profit margins of ISPs can be increased heavily if they encourage more users to switch to such type of system. The load on access links is also reduced by similar proportions, **the island owners have to pay for the Internet access links, on the basis of the usage of such links.** With reduced usage of access links, the Internet consumption bills for island owners can be reduced considerably, which in turn will be a motivation for them to enable IP Multicast support on their networks requiring software (and in some cases hardware) upgrades. Thus, such models is economically sustainable.

Finally, our work is distinct from other similar research because of the following reasons

**Standard compliance:** The proposed method is interoperable with Bit-Torrent protocol. It only requires changes at the end client level, unlike other solutions, which would need network wide support.

**Actual Implementation:** In place of theoretical results or network simulations, we resorted to actually implementing a prototype system of our method and have evaluated it on a large scale real network.

## VII. REFERENCES

[1] B. Cohen, "*Incentives build robustness in Bit-Torrent*", in *Proc. of the First Workshop on the Economics of Peer-to-Peer Systems*, 2003.

[2] S. Saroiu, P. K. Gummadi, and S. D. Gribble, "*A measurement study of peer-to-peer file sharing systems", in Proc. of Multimedia Computing and Networking*, 2002.

[3] N. Leibowitz, A. Bergman, R. Ben-Shaul, and A. Shavit. "*Are file swapping networks cacheable? Characterizing p2p traffic"*, in *Proc. of the 7th Int. WWW Caching Workshop*, 2002.

[4] Amin Vahdat, Ken Yocum, Kevin Walsh, Priya Mahadevan, Dejan Kostic, Je Chase, and David Becker. "*Scalability and accuracy in a large-scale network emulator",* in *Proc. of OSDI*, 2002

[5] Ellen W. Zegura, Ken Calvert, and S. Bhattacharjee. "*How to model an internetwork"*, in *Proc. of IEEE Infocom*, 1996.

[6] Alberto Medina, Anukool Lakhina, Ibrahim Matta, and John Byers. "*Brite: An approach to universal topology generation*", in *Proc. of the International Workshop on Modeling, Analysis and Simulation of Computer and Telecommunications Systems - MASCOTS*, 2001.

[7] Pragmatic General Multicast http://en.wikipedia.org/wiki/Pragmatic_General_Multicast

[8] Network Congestion http://en.wikipedia.org/wiki/Network_congestion

[9] Network Contention http://webopedia.internet.com/TERM/C/contention.html

[10] Bit-torrent used to update workstations. http://torrentfreak.com/university-usesutorrent-080306

[11] Distance vector multicast routing protocol. http://www.ietf.org/rfc/rfc1075.txt

[12] Emulab documentation. http://www.emulab.net/doc.php3

[13] Gnu wget. http://www.gnu.org/software/wget

[14] How to set up Linux for multicast routing. http://www.jukie.net/ bart/multicast/Linux-MroutedMiniHOWTO.html

[15] Ip multicast. http://www.cisco.com/en/US/docs/internetworking/technology/handbook/IPMulti.html

[16] Rfc 3170 - ip multicast applications: Challenges and solutions. http://www.faqs.org/rfcs/rfc3170.html

[17] Pareto Efficiency http://en.wikipedia.org/wiki/Pareto_efficiency

[18] Bit-torrent specification http://wiki.theory.org/BitTorrentSpecification

[19] Bit-torrent location-aware protocol 1.0 specification. http://wiki.theory.org/BitTorrent Location-aware Protocol 1.0 Specification

[20] Dummynet http://info.iet.unipi.it/~luigi/dummynet

[21] The network simulator - ns-2. http://www.isi.edu/nsnam/ns

[22] 3-way Handshake http://www.inetdaemon.com/tutorials/internet/tcp/3-way_handshake.html

[23] Routing information protocol. http://www.ietf.org/rfc/rfc1058.txt